\documentclass[prb, twocolumn, letter, showpacs, floatfix]{revtex4}

\usepackage{graphicx}
\usepackage{calc}
\usepackage{color}

\newlength{\halfwidth}
\newlength{\quarterwidth}
\begin{document}

\title{Flux Dendrites of Opposite Polarity in Superconducting MgB$_2$ rings observed with magneto-optical imaging}
\date{\today}

\author{{\AA}ge Andreas Falnes Olsen}
\email{a.a.f.olsen@fys.uio.no}
\author{Tom Henning Johansen}
\email{t.h.johansen@fys.uio.no}
\author{Daniel Shantsev}
\affiliation{Department of Physics}
\affiliation{Center for Advanced Materials and Nanotechnology, University of Oslo, P.O. Box 1048 Blindern, N-0316 Oslo, Norway}
\author{Eun-Mi Choi}
\author{Hyun-Sook Lee}
\author{Hyun Jung Kim}
\affiliation{National Creative Research Initiative Center for Superconductivity, Department of Physics, Pohang University of Science and Technology, Pohang 790-784, Republic of Korea}
\author{Sung-Ik Lee}
\affiliation{National Creative Research Initiative Center for Superconductivity, Department of Physics, Pohang University of Science and Technology, Pohang 790-784, Republic of Korea}
\affiliation{Quantum Material's Research Laboratory, Korea Basic Science Institute, Daejeon 305-333, Korea}

\begin{abstract}
Magneto-optical imaging was used to observe  flux dendrites with opposite polarities simultaneously penetrate superconducting, ring-shaped MgB$_2$ films. By applying a perpendicular magnetic field, branching dendritic structures nucleate at the outer edge and abruptly propagate deep into the rings. When these structures reach close to the inner edge, where  flux with opposite polarity has penetrated the superconductor,  they occasionally trigger anti-flux dendrites. These anti-dendrites do not branch, but instead trace the triggering dendrite in the backward direction. Two trigger mechanisms, a non-local magnetic and a local thermal, are considered as possible explanations for this unexpected behaviour. Increasing the applied field further, the rings are perforated by dendrites which carry flux to the center hole. Repeated perforations lead to a reversed field profile and new features of dendrite activity when the applied field is subsequently reduced.
\end{abstract}

\pacs{74.70.Ad, 74.25.Qt, 74.25.Ha, 74.78.Db, 68.60.Dv}

\keywords{Superconductivity, dendrites, ring, magnetic instability}

\maketitle

\section{Introduction}

In recent years there has been a growing interest in flux instabilities and catastrophic flux penetration events in superconductors \cite{Altshuler2004}. In particular, one finds that in many superconductor films flux may enter abruptly in the form of magnetic dendrites. While the phenomenon has been observed in  various materials such as  Nb, Nb$_3$Sn, YNi$_2$B$_2$C, NbN, YBa$_2$Cu$_3$O$_x$ (induced by laser pulses), and  patterned Pb films \cite{Duran1995,Rudnev2003,Wimbush2004,Rudnev2005,Leiderer1993,Menghini2005}, it has been most widely studied \cite{Johansen2002,Shantsev2003,Barkov2003,Johansen2001,Bobyl2002,Baziljevich2002,Choi2005,Ye2004,Albrecht2005,Laviano2004} in MgB$_2$. This interest stems in part from the fact that dendrites  are omnipresent in MgB$_2$ films below 10 K and do not need triggering or patterning to occur, and in part from their debilitating effect on the critical current of this material \cite{Zhao2002}, otherwise very promising for many applications \cite{Gurevich2004,Eom2001,Kim2001}. 

It is now generally believed that the dendrites occur as a result of a thermo-magnetic instability, whereby i) motion of vortices releases energy and leads to local heating, and ii) increased temperature leads to a local decrease of the pinning force, enabling enhanced vortex motion. If the released heat is not carried away fast enough, this constitutes a feedback mechanism which induces a thermo-magnetic runaway. Recent experimental work \cite{Ye2004,Albrecht2005,Choi2005, Baziljevich2002} on MgB$_2$ has indeed suggested that such a thermo-magnetic mechanism is a feasible explanation for dendritic instabilities. Lending further support to this picture, it has been shown theoretically  \cite{Aranson2005,Rakhmanov2004, Denisov2006} that the instability will develop into a highly non-uniform pattern if the thermal diffusivity in the superconductor is much smaller than the magnetic diffusivity. Finally, the predictions of these models for the threshold instability field were recently \cite{Denisov2006PRL} found to quantitatively agree  with  experiments on MgB$_2$ and Nb films.

But even as the fundamental mechanism seems to be understood, there are many open questions regarding details in dendritic nucleation and evolution. One of them is the interplay between dendrites of flux and anti-flux. While it was previously  shown that coexisting flux and anti-flux helps the nucleation of dendritic avalanches, \cite{Duran1995,Wimbush2004,Rudnev2005} it has  never been observed how dendrites of opposite polarity interact when they both penetrate a virgin sample.

One geometry where such a situation may be realised is that of a planar ring. \cite{Brandt1997,Navau2005,Pannetier2001} In  zero-field cooled (zfc),  circular superconductor rings  exposed to  perpendicular fields, shielding currents flow around the ring in the same direction everywhere \cite{Brandt1997}. These currents lead to an enhanced field at the outer edge, and a field of opposite polarity at the inner edge.     

In this paper we present results of a magneto-optical (MO) investigation of thin film MgB$_2$ rings showing a rich variety of dendrite behaviour.   The paper is structured as follows: The experimental details are described in section \ref{exp}. Section \ref{results} presents our results, with a discussion of our observations given in section \ref{discussion}.

\section{Experimental details}
\label{exp}

MgB$_2$ films were grown by pulsed laser deposition  on  sapphire substrates. Details on sample fabrication   can be found elsewhere \cite{Kang2001,Choi2005}. Using photolithography, two films, 500~nm thick,  were patterned into circular rings of different size. The lateral dimensions of the larger sample were $r_{outer} = 5$ mm and $r_{inner} = 3$ mm, and the smaller sample  $r_{outer} = 2$ mm and $r_{inner} = 1,2$ mm. 

For observations we used a standard magneto-optical (MO) imaging set-up with a Leica polarisation microscope \cite{Johansen1996}, a liquid helium flow cryostat from Oxford Instruments, a 12-bit Retiga-Exi Fast digital camera from QImaging, and a computer running LabVIEW to aqcuire data and control the applied field. The magnetic sensor was a mirror coated $5 ~\mu$m thick Faraday rotating ferrite garnet film  placed directly on top of the sample. To avoid suppression of dendrites by the metallic mirror layer \cite{Albrecht2005,Choi2005, Baziljevich2002} on the indicator films, we used small, insulating  Ugelstad spheres (monodisperse with diameter $3~\mu$m) as spacers between the film and the sample. 

In a polarisation microscope  the image light intensity is described by Malus law

\begin{equation}
  I = I_{0}\sin^{2} (\theta + \alpha) + I_{b}
\end{equation}

where $\theta$ is the local Faraday rotation of the polarisation (the signal), $\alpha$ is an offset angle from exactly crossed polariser and analyser, $I_{b}$ is the residual intensity at full extinction ($\theta + \alpha = 0$) caused by imperfections in the optical components, and $I_{0} + I_{b}$ is the intensity at maximum opening.  Allowing the offset angle $\alpha$ to be a non-zero value (typically a few degrees) brings two important benefits: firstly the image contrast is improved, and secondly we can distinguish between opposite field directions. In our images bright pixels correspond to positive field, while negative field show up as dark pixels.  In the present report  we have estimated the field-vs-intensity relation using image pixels away from the sample.

The experiments  consisted of ramping the applied field slowly to a maximum level, and then slowly back to zero. The applied field was controlled by computer with a ramp rate   of 0.1~mT/s. Images were recorded  at frequent and regular intervals during the ramp. 

\section{Results}
\label{results}

\begin{figure}
  \includegraphics*[width=\halfwidth, trim=0 0 102 0]{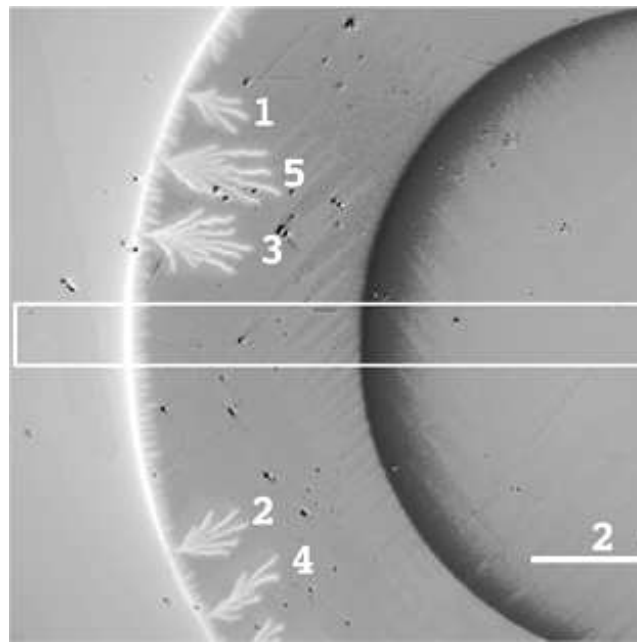} 
  \includegraphics*[width=\halfwidth, trim=0 0 0 0]{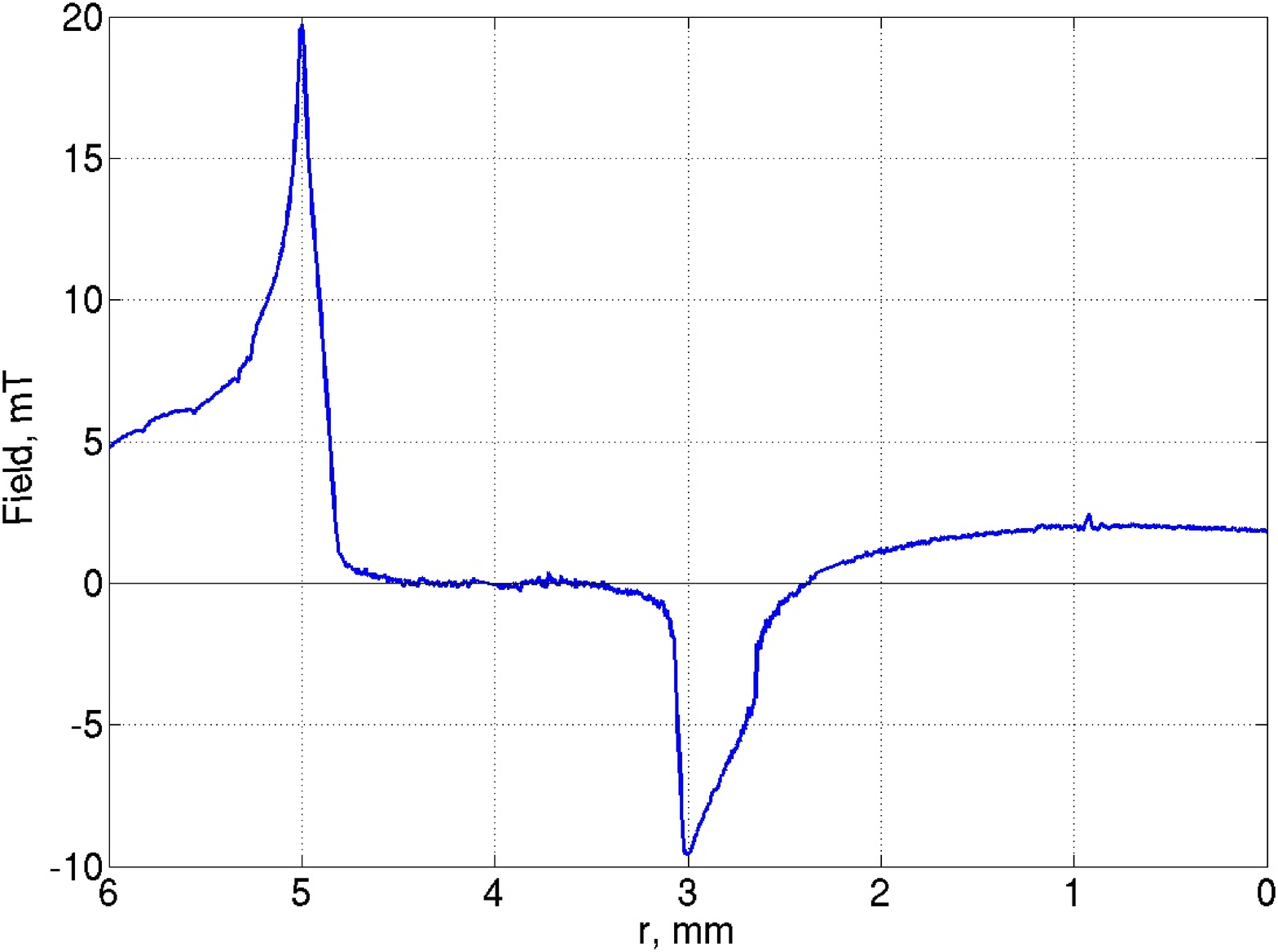}  
  \caption{The flux distribution in the large ring at $B_a = 3.6~$mT. The numbers next to dendrites indicate the order in which they appeared. The profile plot is obtained by averaging vertically within the rectangle. The kink in the profile inside the central hole is an artefact caused by the presence of zig-zag domains in the MO indicator film.}
  \label{critical_state}
\end{figure}

Figure \ref{critical_state} shows the flux distribution in the large ring, initially in the zfc state, at an applied field  $B_a = 3.6$~mT.  Also shown is a flux density profile across the ring averaged over the rectangle indicated in the MO image. At this moderate applied field  the profile shows  smooth flux penetration from the edges in agreement with critical state calculations \cite{Brandt1997} and previous experiments \cite{Pannetier2001} on YBa$_2$Cu$_3$O$_x$. Notice the large positive field at the outer edge and the smaller negative field   at the inner edge.

In the MO image in figure \ref{critical_state} we also see several tree-like  flux structures. Each distinct tree has grown extremely fast. They have been labelled with a number indicating the order in which they appeared. The first of them, labelled 1, appeared  at 3.4~mT. It is seen  how the dendrites become larger with increasing  applied field. However, at  these low fields all of them terminate far from the inner rim, where no activity is seen  apart from a steadily increasing negative field. 

\begin{figure}
  \includegraphics*[width=\halfwidth, trim=0 0 0 0]{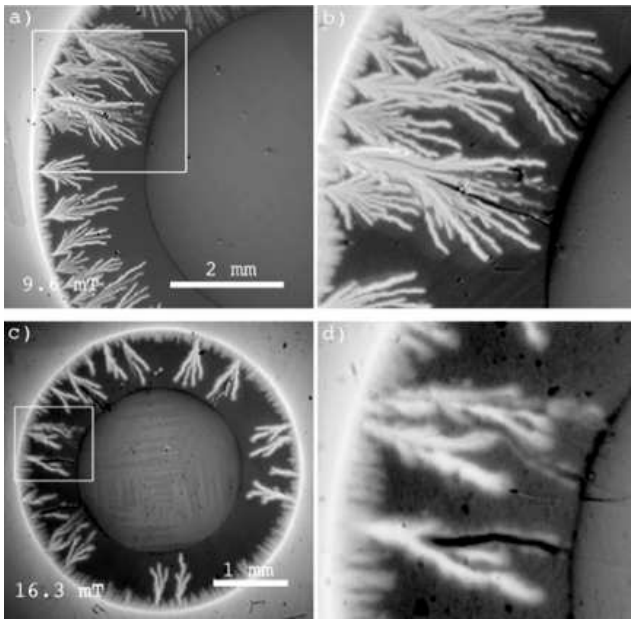} 
  \caption{MO images on zfc samples in increasing applied field. The large ring is shown in a), the small ring in c). In both images the anti-dendrites nucleate near a bright tip, in most cases tracing a bright finger deep into the superconductor. The areas within the rectangles are shown in more detail in b) and d).}
  \label{field_cycle_up}
\end{figure}

The first dendritic structures to almost reach the inner edge appear when the applied field reaches 7~mT. Increasing the field further, anti-flux dendrites appearing as dark fingers, eventually nucleate at the inner edge, see figure \ref{field_cycle_up} a)  where $B_a = 9.6$~mT. The zoomed view in b) shows the details surrounding the two dominating bright structures. Most importantly, the  anti-flux dendrites all originate at a point close to a bright finger tip.  The two large bright dendrites grew at different times, but in both cases the associated dark dendrites appeared in the same image in the sequence. While it seems clear that the anti-dendrites have grown after the dendrites, the two events take place in a very short timespan.

Images c) and d) of figure \ref{field_cycle_up} show  the flux distribution in the small ring at 16.3~mT. The overall features are essentially the same.  In the images one can see a  few dark dendrites that have grown from the inner rim, tracking the core  of some bright dendrites.  Again associated dark and bright dendrites grow simultaneously within our temporal resolution. In fact, it is a general feature of all our experiments that the anti-dendrites  occur in conjunction with a bright dendrite - they coincide both temporally and spatially.

\begin{figure}
  \includegraphics*[width=\halfwidth, trim=0 0 0 0]{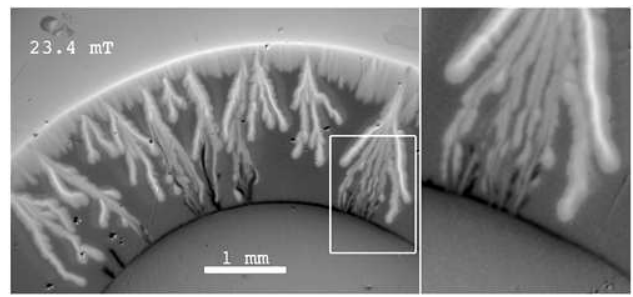} 
  \caption{MO image on fc  samples in increasing applied field. The fc field was 15 mT.  A new detail not seen in the zfc experiment in figure \ref{field_cycle_up} is shown in the zoomed-in view of the  image. Dark and bright dendrites are weaved together in what appears to have been multiple avalanche events. The image has been background corrected by subtracting an image acquired on the virgin sample at 15 mT.}
  \label{field_cycle_up_FC}
\end{figure}

Furthermore, we  observe that while the bright dendrites branch multiple times, the anti-dendrites always consist of just one long finger.   In addition, the dark dendrites usually find a bright branch of a tree from the outer edge and trace that branch quite closely. The image of the small ring in figure \ref{field_cycle_up} d) shows  how close this tracing can be.  

The same observations apply also to field cooled (fc) samples. In figure \ref{field_cycle_up_FC} the images show dark fingers which grow deep into bright trees that originate at the outer edge. Just as for the zfc experiments, the anti-dendrites are temporally and spatially strongly correlated with bright dendrites. An interesting detail can be seen in the zoom view of figure \ref{field_cycle_up_FC}, where dendrites and anti-dendrites are stacked on top of each other as if they were weaved together.

\begin{figure}
  \includegraphics*[width=\halfwidth]{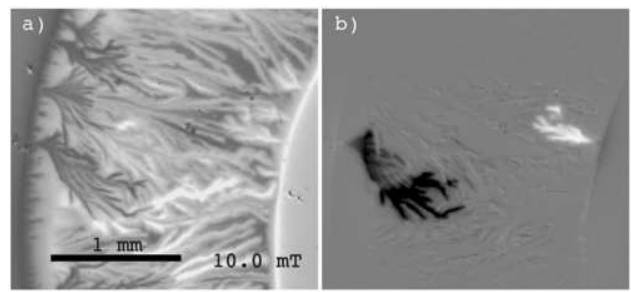}
  \caption{An MO image of a zfc sample during field descent is shown in (a). The image in b) is obtained by subtracting the previous image in the sequence, thus  highlighting the growth of specific dendrites, as well as showing small-scale flux rearrangements in a large region in response to the dendritic avalanches.}
  \label{field_cycle_down}
\end{figure}

\begin{figure*}[!ht]
  \includegraphics*[width=\halfwidth, trim=0 0 0 0]{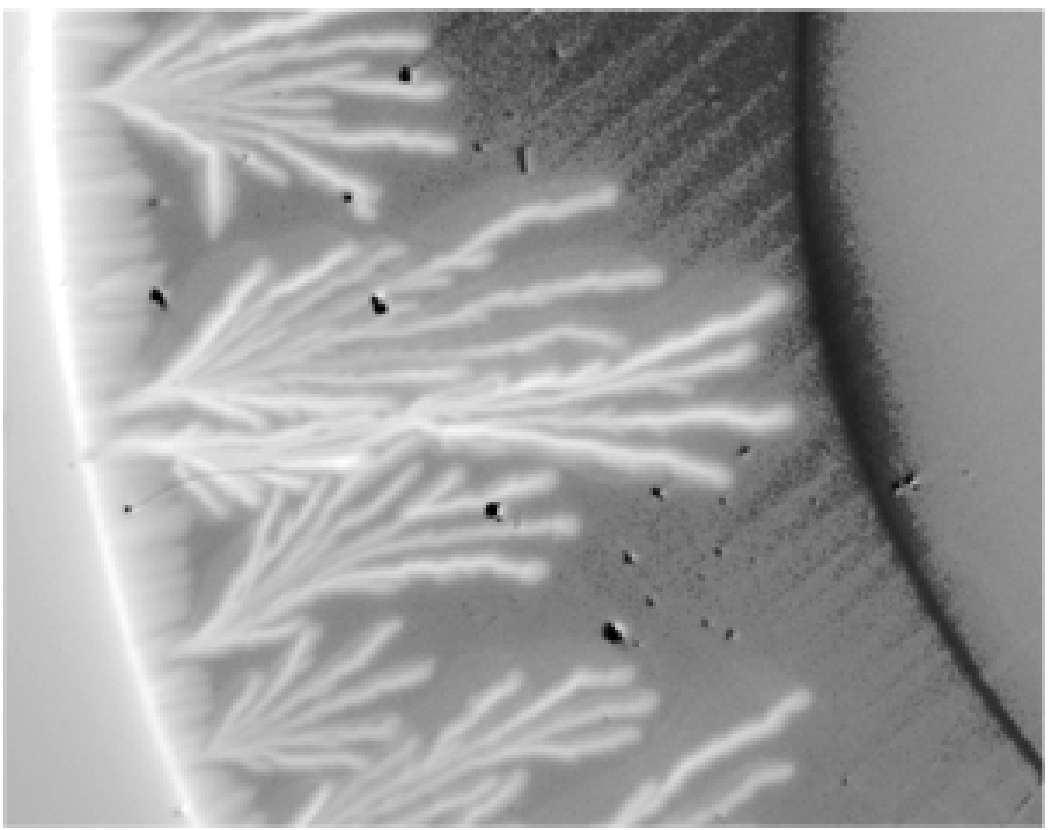} 
  \includegraphics*[width=\halfwidth, trim=0 0 0 0]{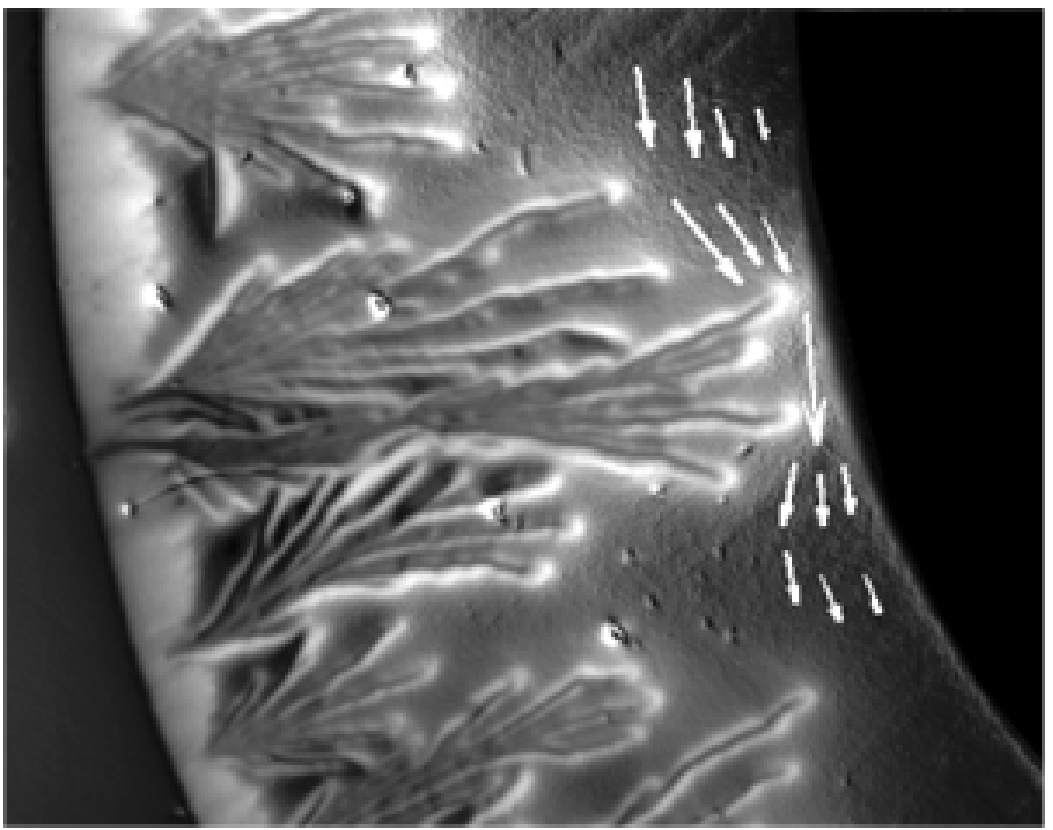} 
  \caption{Left: The MO image shows the field distribution near a bright dendrite which has grown nearly all the way to the center hole. Note the relatively strong negative field at the inner edge close to the finger. Right: A current density map obtained by inverting the B-field image using the Biot Savart law. The arrows show the direction of current. There is an increased current density between the inner edge and the tip of the bright finger. The images suggest that if bright dendrites come close enough to the inner edge, they may trigger growth of a dark dendrite.}
  \label{J_dendrite}
\end{figure*}

Returning  to the zfc experiments, the behaviour is different when we decrease the applied field from its maximum value. The field at the outer and inner edges then decreases and increases, respectively. Flux of opposite polarity - dark on the outer edge, bright on the inner - penetrates the sample, with a regular penetration being interrupted by dendritic structures, see  figure \ref{field_cycle_down}.  In order to illustrate more clearly the dynamical aspects, image b) displays the difference between subsequent images in the sequence.  Where the flux density is unchanged, pixels are gray. Dark pixels indicate that flux has left or  anti-flux has entered. Dendritic structures, dark  from the outer and bright from the inner edges,  have appeared in both images, meaning they  nucleated at the same time. However, unlike the behaviour in increasing field, we find that i) the tips of the structures are far apart, ii) neither of them come close to  the opposite edge, and iii) both are branching. 

\section{Discussion}
\label{discussion}

We have found that the anti-dendrites forming in increasing field  appear to be triggered by bright dendrites approaching the inner rim. There are at least two  possible triggering mechanisms: i) a non-local magnetic coupling where the negative field at the inner edge is enhanced by the sudden appearance of a bright dendrite, and ii) a local thermal mechanism where heat associated with the bright dendrite tip facilitates the nucleation and growth of an anti-dendrite. 

The enhancement of the negative field is demonstrated  in figure \ref{J_dendrite}, showing the field (left) and current (right) maps close to a long dendrite which almost comes across to the center hole. The current map shows a significantly increased current density between the finger tip and the inner edge. One can understand this increase by considering the current flow around dendrites. Figure \ref{J_dendrite} shows that the current density is large along  flux fingers, but very small in the dendrite cores, implying that little current flows across the dendrite. Indeed,   previous work on MgB$_2$ has indicated that the current density is in fact  maximum  along  dendrites, \cite{Laviano2004,Shantsev2003,Barkov2003} thus making them  effective barriers against additional current.  As a result, the Meissner state  currents otherwise flowing throughout the  ring  become  concentrated near the inner edge,  resulting in a regional increase in the magnitude of the  negative edge field. An anti-dendrite can form provided the magnitude exceeds a threshold value, since a   typical feature of   both conventional  \cite{Swartz1968} and dendritic \cite{Barkov2003,Denisov2006} flux jumps is the existence of a threshold field which must  be exceeded for an avalanche to occur. Furthermore, the abrupt character of the field increase will lead to a large electric field  which also helps the nucleation \cite{Denisov2006,Aranson2005} of an anti-dendrite.

In addition to this non-local effect, a bright dendrite that reaches the negative flux region near the inner edge will also induce a local temperature increase. This is because the core of the bright dendrite is itself a region of increased temperature,  and  because the ensuing  flux annihilation releases heat. The resulting  elevated temperature in a small area helps trigger an anti-dendrite, much like the laser pulse triggering  \cite{Leiderer1993} on YBa$_2$Cu$_3$O$_x$. With MO imaging  it is very difficult to determine how far from the inner edge a bright dendrite actually stops when an anti-dendrite has grown on top of it. From our experiments  we are unable to tell whether bright and dark regions have been in contact prior to the nucleation of the anti-dendrite, and thus whether the thermal trigger mechanism is feasible.  This important issue is open for future study using ultra-fast MO imaging techniques \cite{Freeman1992,Bujok1993}.

Once anti-dendrites have been triggered they tend to trace the  bright dendrite whose appearance triggered them. We believe this tracing is assisted by the flux-anti-flux attraction, heating as a result of flux annihilation, and possibly the residual heat in the core of the bright dendrite. These three effects  help contain the anti-dendrite tip within the bright finger, and hence also lead to the observed suppression of branching.

\begin{figure*}
  \includegraphics*[width=\halfwidth, trim=150 350 500 100]{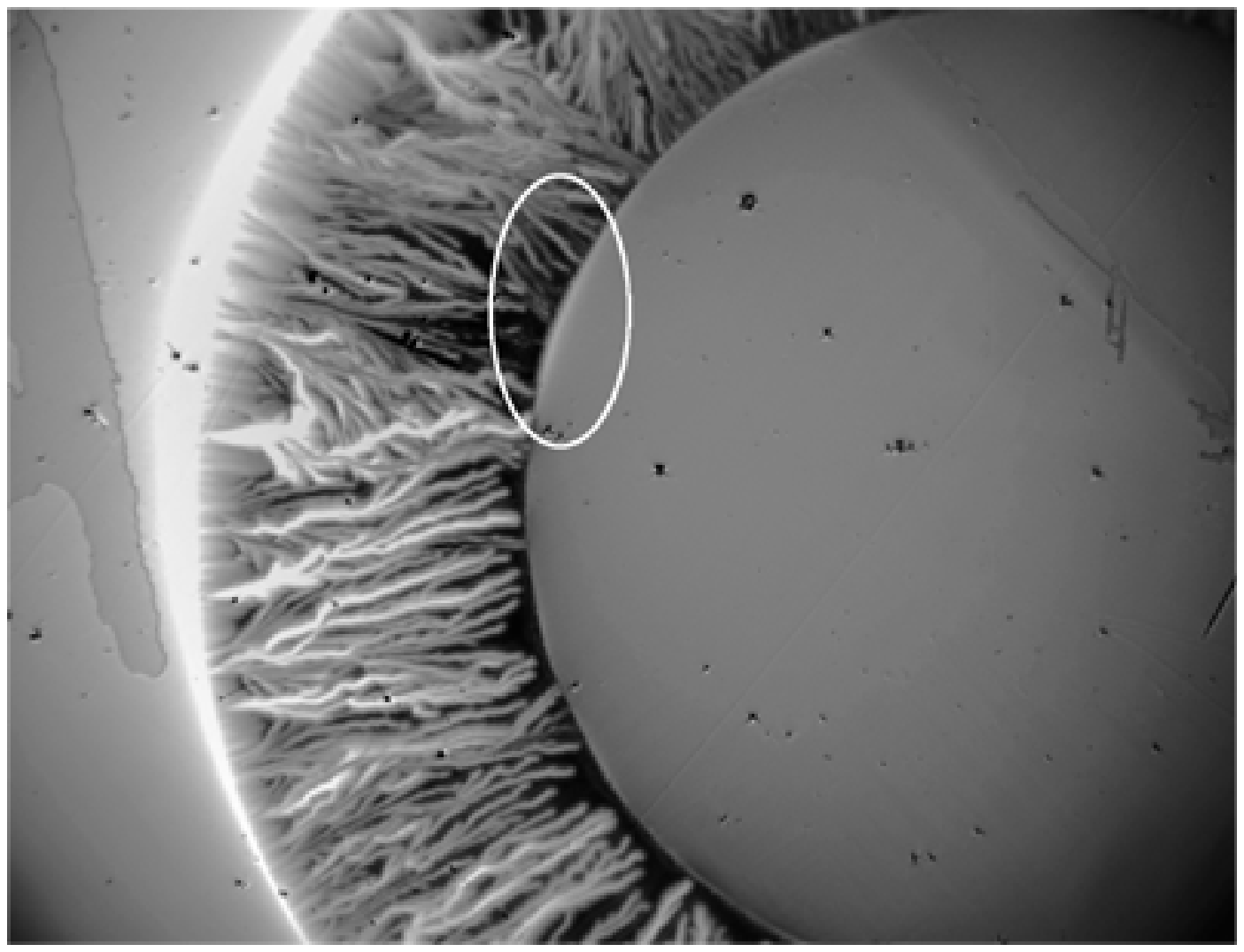} 
  \includegraphics*[width=\halfwidth, trim=150 350 500 100]{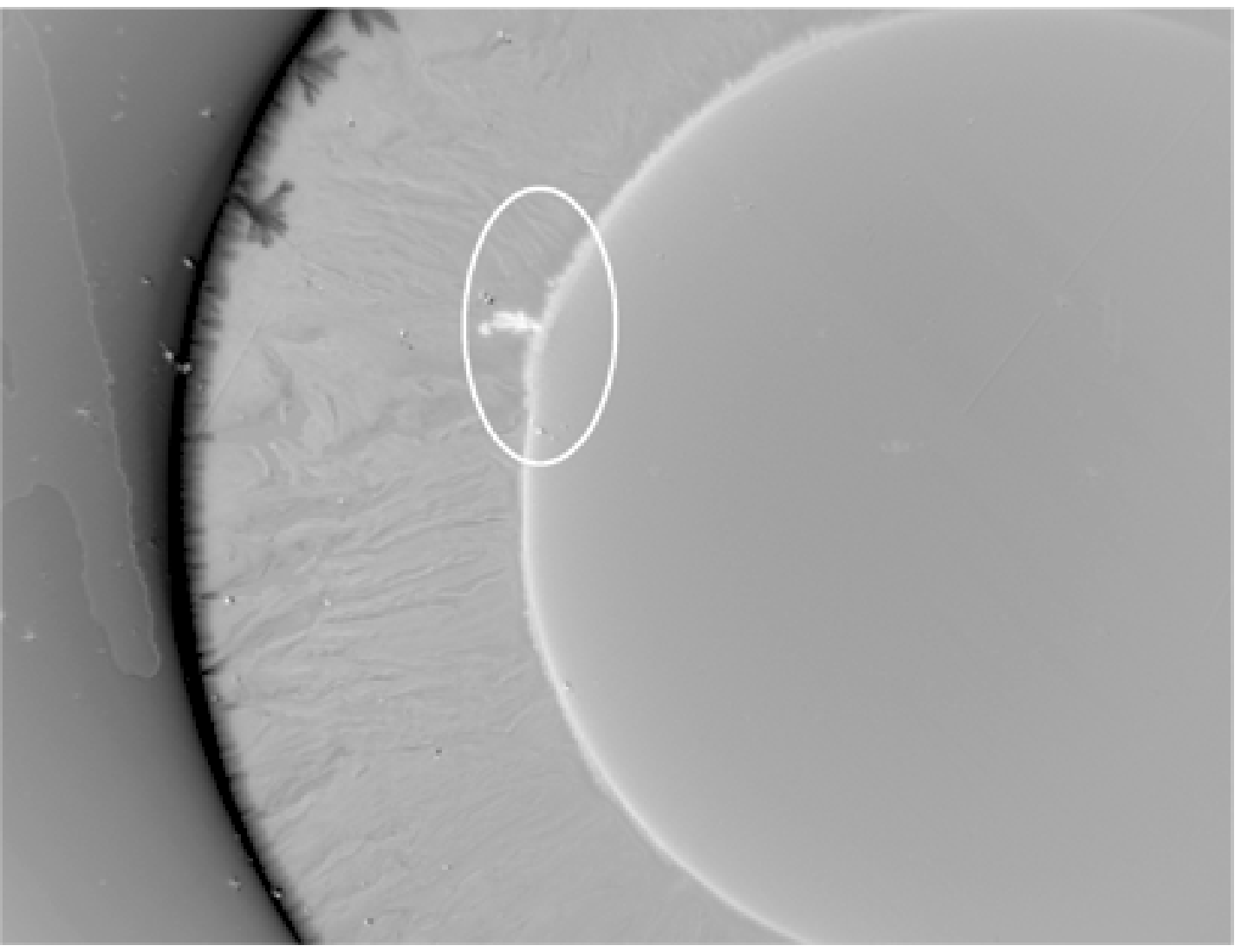}  
  \caption{Left: MO image recorded at maximum applied field of 20~mT. Notice that the field is positive near the inner edge. Right: image recorded at 12,6~mT after subtracting the peak field image. Dark pixels indicate decrease in flux, and bright pixels indicate increase. The difference image resembles what we see in a virgin sample at small applied fields, with a regular flux decrease at the outer edge and flux increase  at the inner edge. The nucleation spot of the first dendrite is a region where we find a large positive edge field at peak $B_a$, while the flux density in the superconductor is quite low.}
  \label{descending}
\end{figure*}

For low applied fields the magnitude of the field at the outer edge is larger than that at the inner edge. In consequence,  the threshold field is reached sooner at the outer edge  and dendrites nucleate there first. This fact   has profound implications for the dynamics at the inner edge. Before the negative field at the inner edge reaches the threshold, bright dendrites perforate the ring, bringing positive flux from the outside to the central hole. After the first perforation event, new events are frequent and increase the average flux density at the inner edge to positive values, so the net effect of increasing the applied field is to increase the field at the inner edge as well.  This is demonstrated in figure \ref{descending}, where the left MO image shows the flux distribution at peak applied field, $B_a = 20$~mT.

While the positive inner edge field explains why no dark denrites form in increasing applied field, one needs to examine the images in figure \ref{descending} more closely to understand why bright dendrites form when $B_a$ is subsequently decreased. Of particular importance are regions  where the field is  large and positive  at the inner edge, while the flux density inside the superconducting material is small, see e.g the encircled area in the images.  In such regions the field gradient inside the superconducting material is the opposite of what one would see in  the case of regular penetration. Moreover, as is shown in the right image in figure \ref{descending}, where $B_a$ has been decreased to 12.6~mT, the flux change is initially uniform at the two edges, meaning that the already positive edge field in the encircled region has increased further. Thus a modest decrease in applied field is sufficient to induce the bright dendrite we see in the right image in figure \ref{descending}. The result is that while the perforations \emph{impede} the nucleation of dark dendrites  in increasing $B_a$, they \emph{facilitate} nucleation of bright dendrites in decreasing $B_a$. 

\section{Conclusions}

Flux dendrites which nucleate at the outer edge of a superconducting MgB$_2$ ring lead to unexpected flux penetration at the inner edge. In particular, we have found that when increasing the applied field to an intermediate level, i)  dendrites and anti-dendrites nucleate at the outer and inner edges of the rings, respectively; however, all anti-flux dendrites are triggered by large flux dendrites; ii) anti-dendrites  do not branch, instead they find a finger of the triggering positive dendrite and trace it closely. The triggering can occur either due to a locally enhanced magnetic or electric field, or due to a local temperature elevation in the negative flux near the inner edge. Ultra-high temporal resolution is needed in order to conclusively decide which of the two mechanisms is dominant. We further found that  iii) for larger applied field very large dendrites perforate the rings,  bring flux into the center hole, and ultimately reverse the field profile near the inner  edge; and iv) the reversed field profile leads to prolific nucleation of flux dendrites at the inner edge when the applied field is subsequently reduced. 

\begin{acknowledgments}
This work was sponsored by FUNMAT@UiO.
\end{acknowledgments}

\end{document}